\RequirePackage{lineno}
\documentclass[prd,twocolumn,showpacs,preprintnumbers,floatfix,amsmath,amssymb,amsfonts]{revtex4}

\usepackage{graphicx}
\usepackage{amsmath}
\usepackage{dcolumn}
\usepackage{epsfig}
\usepackage{psfrag}
\usepackage{subfigure}
\usepackage{hyperref}

\RequirePackage{xspace}

\def\TeV{\ifmmode {\mathrm{\ Te\kern -0.1em V}}\else
	                   \textrm{Te\kern -0.1em V}\fi}%
\def\GeV{\ifmmode {\mathrm{\ Ge\kern -0.1em V}}\else
	                   \textrm{Ge\kern -0.1em V}\fi}%
\def\MeV{\ifmmode {\mathrm{\ Me\kern -0.1em V}}\else
	                   \textrm{Me\kern -0.1em V}\fi}%
\def\keV{\ifmmode {\mathrm{\ ke\kern -0.1em V}}\else
	                   \textrm{ke\kern -0.1em V}\fi}%
\def\eV{\ifmmode  {\mathrm{\ e\kern -0.1em V}}\else
	                   \textrm{e\kern -0.1em V}\fi}%
\let\tev=\TeV
\let\gev=\GeV

\def\TeVc{\ifmmode {\mathrm{\ Te\kern -0.1em V}/c}\else
	                   {\textrm{Te\kern -0.1em V}/$c$}\fi}%
\def\GeVc{\ifmmode {\mathrm{\ Ge\kern -0.1em V}/c}\else
	                   {\textrm{Ge\kern -0.1em V}/$c$}\fi}%
\def\MeVc{\ifmmode {\mathrm{\ Me\kern -0.1em V}/c}\else
	                   {\textrm{Me\kern -0.1em V}/$c$}\fi}%
\def\keVc{\ifmmode {\mathrm{\ ke\kern -0.1em V}/c}\else
	                   {\textrm{ke\kern -0.1em V}/$c$}\fi}%
\def\eVc{\ifmmode  {\mathrm{\ e\kern -0.1em V}/c}\else
	                   {\textrm{e\kern -0.1em V}/$c$}\fi}%
	
\def\cm{\ifmmode  {\mathrm{\ cm}}\else	                   
\textrm{~cm}\fi}%

\def\mm{\ifmmode  {\mathrm{\ mm}}\else	                   
\textrm{~mm}\fi}%
	
\def\ifb{\mbox{fb$^{-1}$}}

\usepackage{relsize}
\def\babar{\mbox{\slshape B\kern-0.1em{\smaller A}\kern-0.1em
    B\kern-0.1em{\smaller A\kern-0.2em R}}}

\begin{document}

\title{New approach to identifying boosted hadronically decaying particles using jet substructure in its  center-of-mass frame}

\author{Chunhui Chen} 
\affiliation{Department of Physics and Astronomy, Iowa State University, Ames, Iowa 50011, United State}
\affiliation{High Energy Physics Division, Argonne National Laboratory, Argonne, IL 60439, United State}

\begin{abstract}
In this paper we introduce a new approach to study jet substructure in the center-of-mass frame of the jet. 
We demonstrate that it can be used to discriminate the boosted heavy particles from the QCD jets 
and the method is complementary to other jet substructure algorithms.  
Applications to searches for hadronically decaying $W/Z$+jets
and heavy resonances that decay to a $WW$ final state are also discussed.
\end{abstract}

\maketitle

\section{Introduction}
\label{sec:intro}
Many theories beyond the standard model (SM) predict new particles 
with masses at the TeV scale. Some of these heavy resonances, such  as a $Z^\prime$, a $W^\prime$, 
a heavy Higgs or fourth generation quarks, can decay to final states with an electroweak gauge boson, 
$W$ or $Z$, or a top quark $t$. Because of the energy scale of these processes,
the $W$, $Z$ and $t$ from the heavy resonance decay are
highly boosted. Their hadronically decaying  products are often so 
collimated that they appear as a single jet, hereafter called $W$, $Z$ or $t$ jets. 
The presence of boosted $W$, $Z$ and $t$ jets gives us a unique experimental
signature to look for new physics (NP) phenomena beyond the SM. 

In recent years, theoretical and experimental studies have been performed to investigate the signature 
of boosted particles, not only including 
$W$ and $Z$ bosons~\cite{Butterworth:2002tt,Almeida:2008yp,Ellis:2009su,Ellis:2009me,Hackstein:2010wk,Katz:2010mr,Thaler:2010tr,Cui:2010km} and 
$t$ quarks~\cite{Thaler:2008ju,Kaplan:2008ie,Almeida:2008tp,Krohn:2009wm,Plehn:2010st,Chekanov:2010vc,Bhattacherjee:2010za,Rehermann:2010vq,Chekanov:2010gv}
but also a boosted light 
Higgs boson~\cite{Butterworth:2008iy,Plehn:2009rk,Kribs:2009yh,Soper:2010xk,Chen:2010wk, Falkowski:2010hi,Kribs:2010hp,Almeida:2010pa,Katz:2010iq,Kim:2010uj}
at the LHC. In these studies, the complete final state of the heavy particle is reconstructed as a single jet. The invariant mass
of the reconstructed jet ($m_{\rm jet}$) is therefore a good indicator of its origin.
It has been shown that by using the technique
of boosted jets, one can often achieve comparable, and sometimes even better sensitivities to
probe NP at the TeV scale. However, one experimental challenge in the application of boosted 
$W$, $Z$ and $t$ jets is the copious production of QCD jets
at the LHC, where the QCD jets are defined as those jets initiated by a non-top quark or gluon.
As a result,  the jet mass alone may not provide sufficient discriminating power to
effectively distinguish $W$, $Z$ and $t$ jets from the overwhelming QCD background in many analyses. 
In the last few years many techniques have been developed to address this issue by  exploring jet substructure
as an additional experimental handle to  identify boosted heavy objects.

In general, jet substructure techniques can be classified into two categories. The first category employs jet shape
observables~\cite{Abdesselam:2010pt} to probe the energy distributions inside jets. The second category uses jet-grooming algorithms, including
filtering~\cite{Butterworth:2008iy}, pruning~\cite{Ellis:2009su,Ellis:2009me} and trimming~\cite{Krohn:2009th}. 
They take advantage of the characteristics of the subjets within a jet by 
reclustering the energy clusters of a jet with the $k_T$ or Cambridge-Aachen (CA) sequential jet reconstruction algorithms.
So far, most  jet substructure techniques are based on energy clusters  measured in the lab frame.
In this paper, we introduce a different approach to study  jet substructure in the center-of-mass frame of the jet.
A similar idea has also been explored to search for hadronically decaying Higgs boson~\cite{Kim:2010uj}.

We organize this paper as follows: In Section.~\ref{sec:sample}, we describe the event sample we used  in the study.
Section~\ref{sec:jet_sub} discusses the method to study jet substructure in the jet center-of-mass frame and its performance.
Several example of the application of our method are given in Section~\ref{sec:app}.
We conclude in Section~\ref{sec:conclusion}.


\section{Event Sample}
\label{sec:sample}
We use boosted $W$ jets, from the SM process of 
$W$+jets production,  as an example to illustrate our proposed jet substructure method. 
For simplicity we only consider the background from the SM dijet production since its cross section 
is several orders of magnitudes larger than the other SM processes. However, our method
is generic and is applicable to all boosted hadronically decaying objects, such as the
$Z$ boson, Higgs boson, or $t$ quark. In addition, we also generate events to simulate the SM $Z$+jets production 
and a heavy-particle $X$ that decays to a $WW$ final state.

All the events used in this analysis are produced using the P{\footnotesize ythia} 6.421 event generator~\cite{Sjostrand:2006za}
for the $pp$ collision at $7\,\rm TeV$ center-of-mass energy.
The P{\footnotesize ythia} parameters are set to the default ATLAS parameters tuned to describe expected multiple interactions.
In order to simulate the finite resolution of the
Calorimeter detector at the LHC, we divide the $(\eta, \phi)$ plane into $0.1\times 0.1$ cells. We sum over the energy
of particles entering each cell in each event, except for the neutrinos and muons, and replace it with a massless pseudoparticle
of the same energy, also referred as an energy cluster,  pointing to the center of the  cell. These pseudoparticles are fed into the 
F{\footnotesize astJet}~\cite{fastjet} package for  jet reconstruction.
The jets are reconstructed with the anti-$k_T$ algorithm~\cite{Cacciari:2008gp}  
with a distance parameter of $\Delta R=0.6$. Currently the  anti-$k_T$ jet algorithm is the default one used at the ATLAS and CMS experiments. 

\section{Jet Substructure in the rest frame}
\label{sec:jet_sub}
In this section we describe the method to study jet substructure in the center-of-mass frame of
the jet in order to distinguish the boosted hadronically decaying
particle from the QCD jets. 
We select jets with  $p_{\rm T}\ge\,300\,\gev$ and $|\eta|\le2.5$ as $W$ jet candidates.
We further require that the $W$ jet candidates have $40\,\gev\le m_{\rm jet}\le 140\,\gev$. In case there is more than one candidate in
an event, we keep the $W$ jet candidate with the highest $p_{\rm T}$ in the event.  
Studies using $W$+jets Monte Carlo (MC) samples show that this procedure results in the selection of the correct $W$ jet signal candidate more than 90\% of the time.

\subsection{Center-of-mass  frame of  a jet}
\begin{figure}[!htb]
\begin{center}
\includegraphics[width=0.45\textwidth]{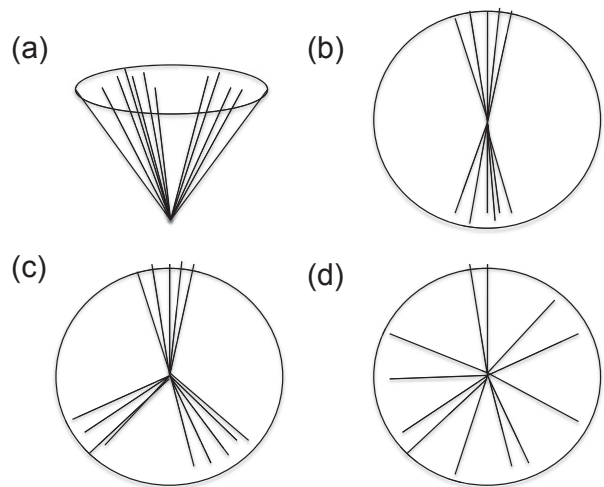}
\caption{Illustration of the constituent particle distribution of a jet. (a) Jet in the lab frame. (b) Jet of a boosted particle decaying to a two-body 
final state in the jet rest  frame. (c) Jet of a boosted particle decaying to a three-body 
final state in the jet rest frame. (d) QCD jet in its rest frame. }
\label{fig:Bst_scheme}
\end{center}
\end{figure}

We define the center-of-mass frame (rest frame) of a jet as the frame where the four momentum of the
jet is equal to $p^{\rm rest}_{\mu}\equiv (m_{\rm jet}, 0, 0, 0)$. A jet consists of its constituent particles.
The distribution of the constituent particles of a boosted $W/Z$, $t$ or Higgs jet in its center-of-mass frame,
is almost identical to those of the $W/Z$, $t$ or Higgs particle produced at rest. For example, in the
rest frame of a hadronically decaying $W$ boson, the constituent particles look like a back-to-back di-jet
event. Similarly for the hadronically decaying $t$ quark, its constituent  particle
distribution has a three body decay topology in its rest frame. On the other hand, a QCD jet acquires its mass through
gluon radiation and it is not a closed system. Its constituent particle distribution in the rest frame does not correspond 
to any physical state and is more likely to be random, as illustrated in Figure~\ref{fig:Bst_scheme}. This observation is in analogy
 to the one in the $e^+e^-\to\Upsilon(4S)$ experiments, such as \babar\  and Belle. In the latter case  the event shape is used to help disentangle
the $B\bar{B}$ signal, whose decay products have an isotropic distribution, from the continuum background that has a pronounced
two-jet  structure. As a result, by going to the jet rest frame, we can apply the knowledge of the event shape variables
learned from  $e^+e^-$ experiments to the experiments at  the LHC in order to separate the boosted heavy objects from the QCD
jets . Furthermore, the correlation between the jet substructure and jet momentum is expected to be small by definition.

\subsection{Shape variables}
\begin{figure*}[p]
\begin{center}
\includegraphics[width=0.38\textwidth]{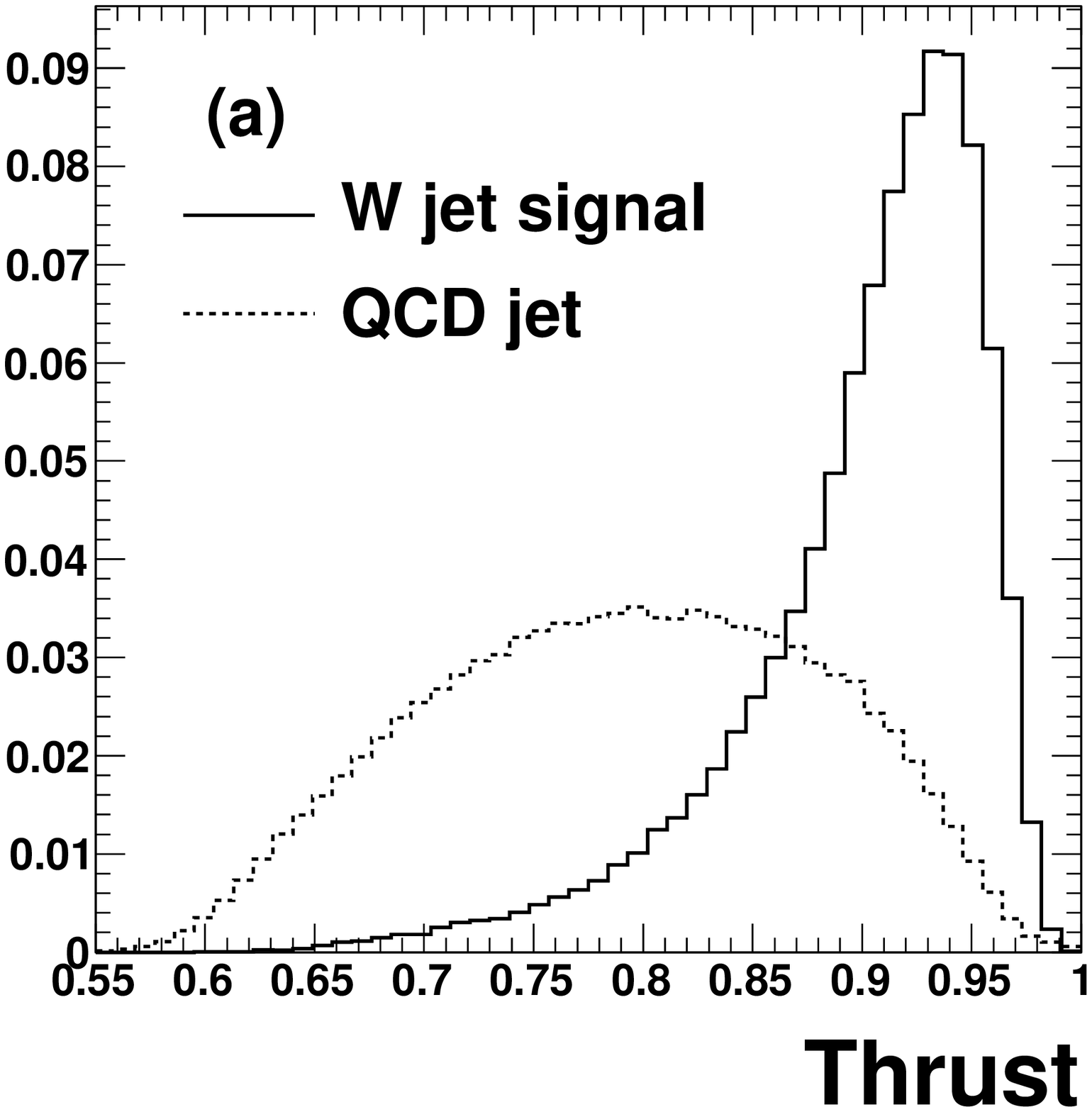}
\includegraphics[width=0.38\textwidth]{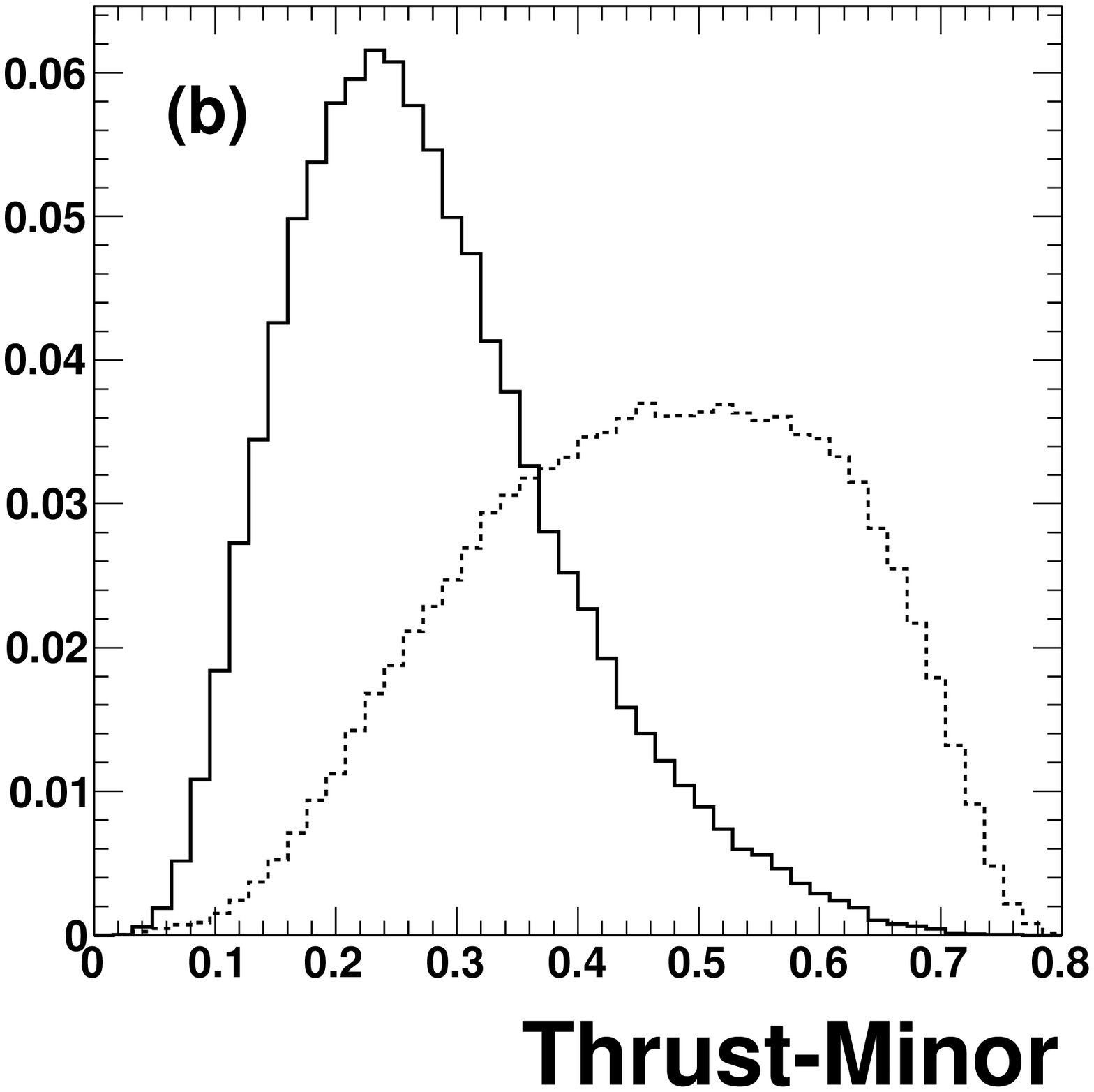}
\includegraphics[width=0.38\textwidth]{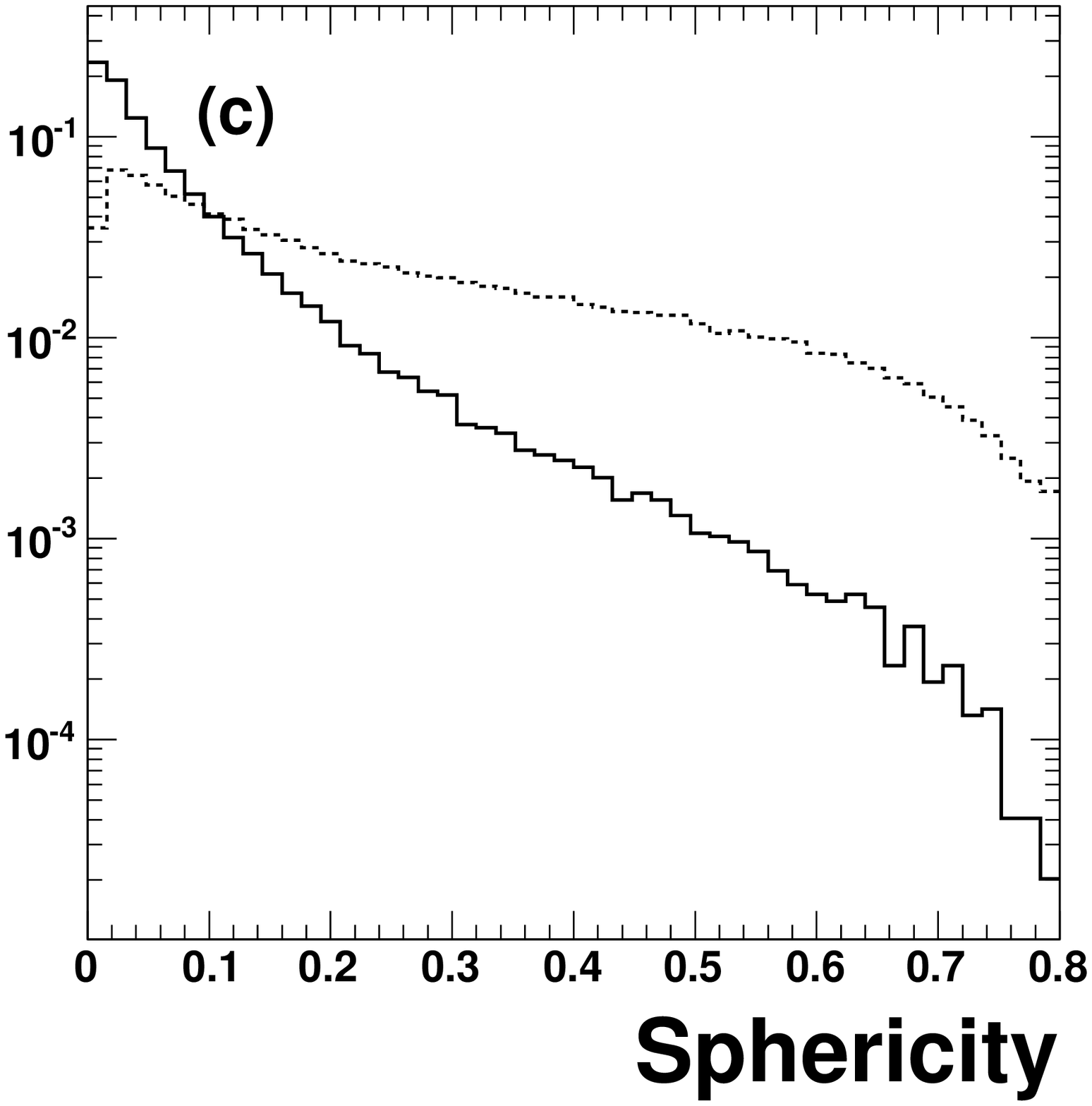}
\includegraphics[width=0.38\textwidth]{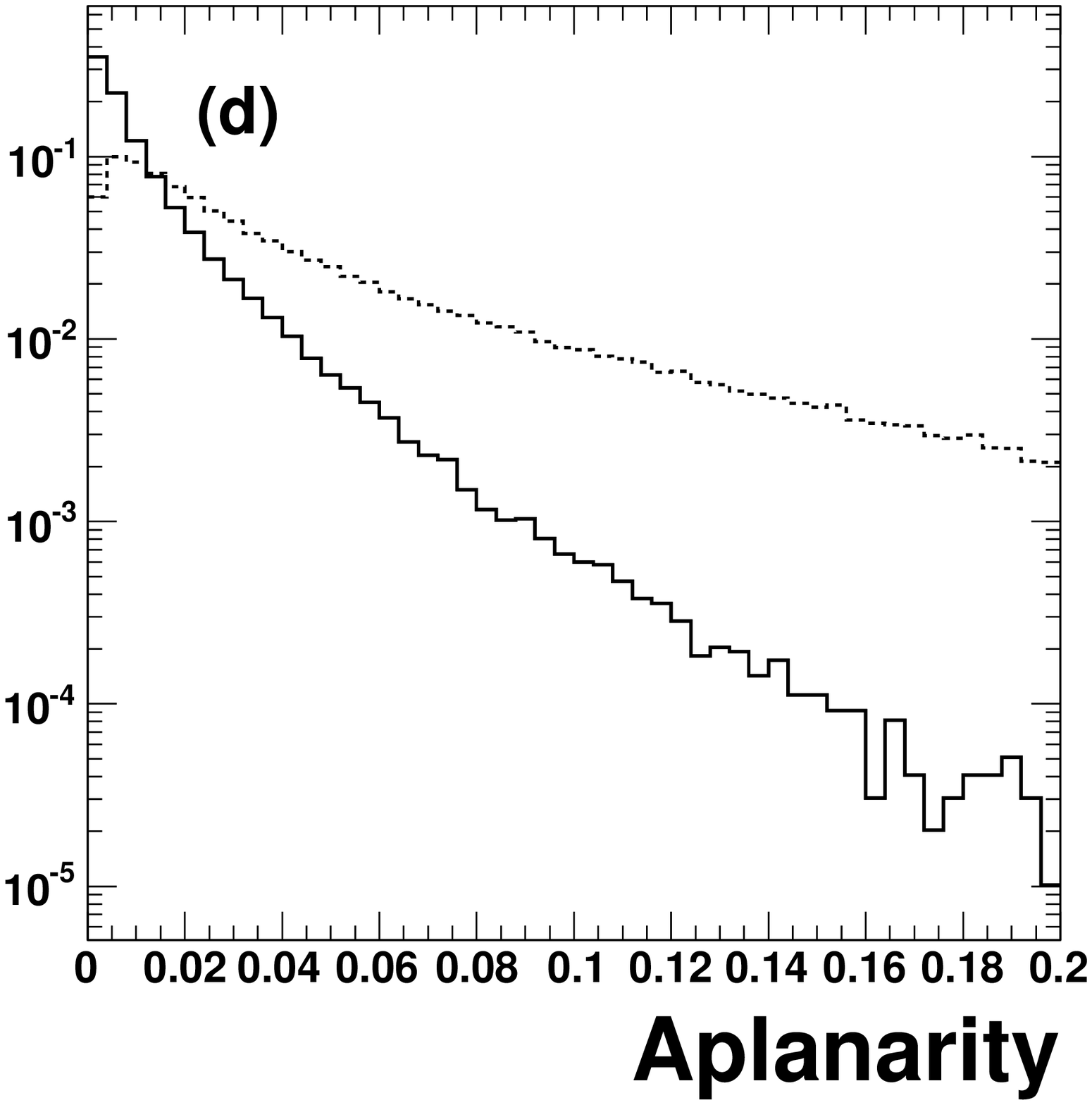}
\includegraphics[width=0.38\textwidth]{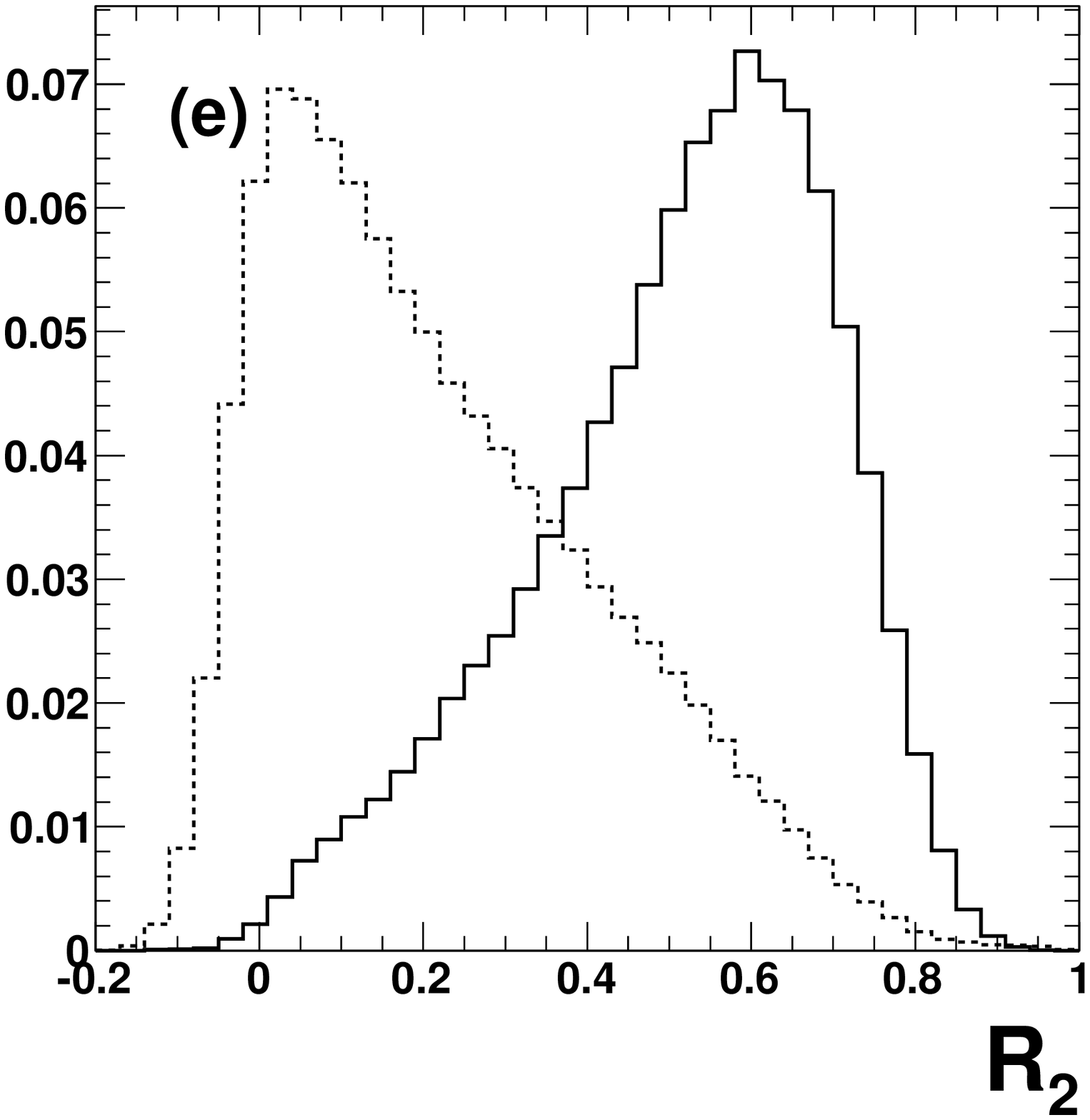}
\includegraphics[width=0.38\textwidth]{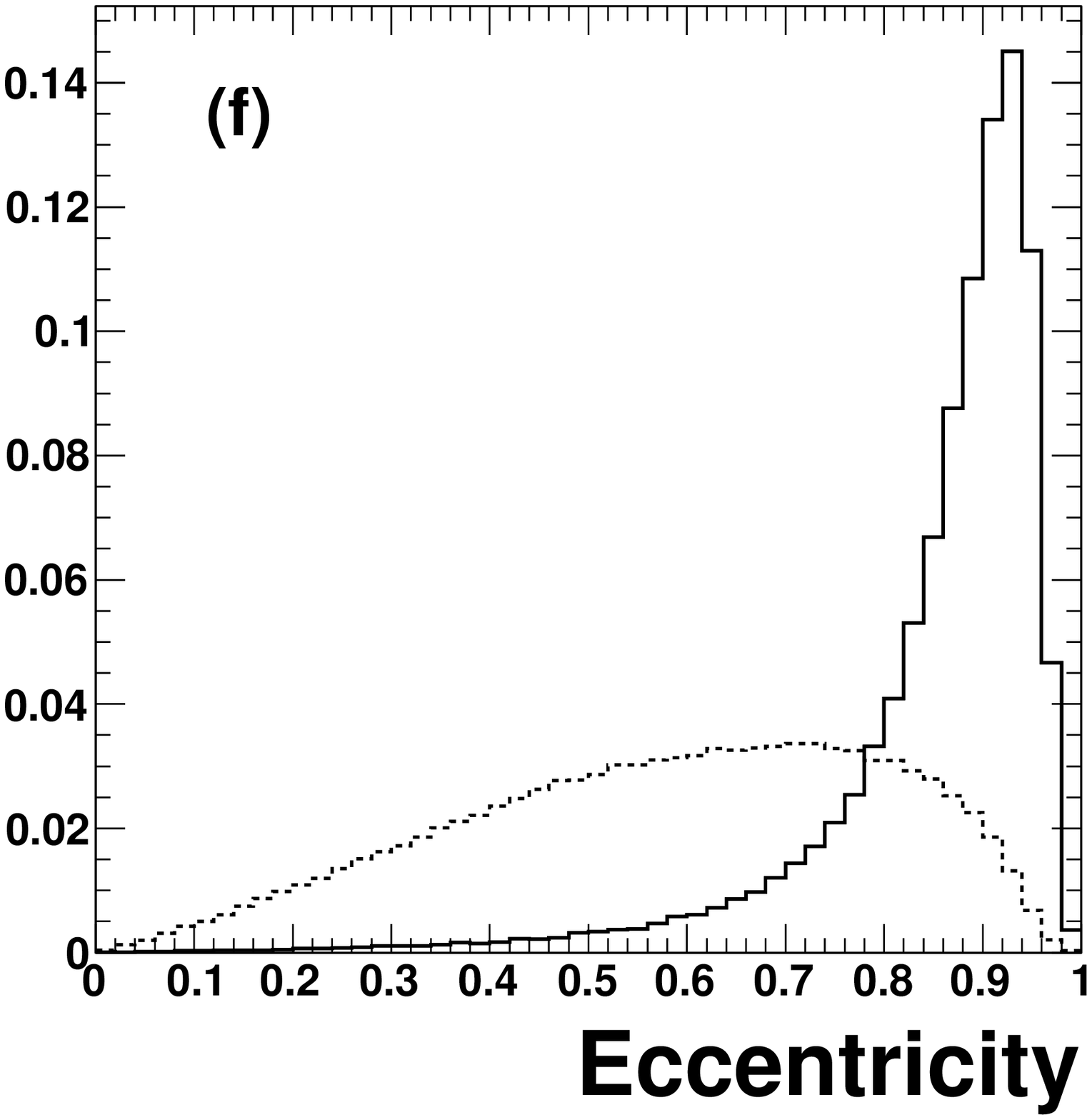}
\caption{The distributions of the jet shape variables: (a) Thrust, (b) Thrust-minor, (c) Sphericity,
(d) Aplanarity, (e) $R_2$ and (f) Eccentricity for the $W$ jet signal and QCD jet background. The eccentricity is a commonly used 
shape variable in the lab frame and is shown here for the purpose of comparison. All the distributions are normalized to unity.}
\label{fig:shape}
\end{center}
\end{figure*}

We introduce five shape variables that are commonly used at the $e^+e^-$ experiments~\cite{babar_book}.
All the variables are calculated using the energy clusters of a jet in its center-of-mass frame and they are:
\begin{itemize}
\item Thrust:
The thrust axis~\cite{Brandt:1964sa,Farhi:1977sg} of a jet in its center-of-mass frame, $\hat{T}$, is defined as the direction which maximizes the
sum of the longitudinal momenta of the energy clusters. The thrust, $T$, is related to this direction and is calculated as:
\begin{equation}
T=\frac{\sum_i |\hat{T}\cdot\vec{p}_i|}{\sum_i|\vec{p}_i|},
\end{equation}
where  $\vec{p}_i$ is the momentum of each  energy cluster in the jet rest frame. The allowed range of 
$T$ is between 0.5 and 1, where $T=1$ corresponds to a highly directional distribution of the energy clusters,
and $T= 0.5$ corresponds to an isotropic distribution.
\item Thrust minor: The thrust minor~\cite{Brandt:1964sa,Farhi:1977sg}, $T_{\rm min}$, is defined as:
\begin{equation}
T_{\rm min}=\frac{\sum_i |\vec{p}_i\times \hat{T}|}{\sum_i |\vec{p}_i|}.
\end{equation}
$T_{\rm min}=0$ corresponds to a highly directional distribution of the energy clusters,
and $T= 0.5$ corresponds to an isotropic distribution.
\item Sphericity: The sphericity tensor~\cite{Bjorken:1969wi} is defined as:
\begin{equation}
S^{\alpha,\beta}=\frac{\sum_i p^\alpha_i p^\beta_i}{\sum_i |\vec{p}_i|^2},
\end{equation}
where $\alpha$ and $\beta$ correspond to the $x$, $y$ and $z$ components  of the 
momentum of each energy cluster in the jet rest frame. By standard diagonalization of $S^{\alpha \beta}$ one 
may find three eigenvalues $\lambda_1 \geq \lambda_2 \geq \lambda_3$, with $\lambda_1 + \lambda_2 + \lambda_3 = 1$. 
The sphericity is then defined as \
\begin{equation} S = \frac{3}{2} \, (\lambda_2 + \lambda_3).
\end{equation}
Sphericity is a measure of the summed  squares of transverse momenta of all the energy clusters with respect to the jet axis, and $0\le S \le1$.
A jet with two back-to-back subjets in its rest frame has $S=0$, and $S=1$ indicates an isotropic distribution of the
energy clusters.
\item Aplanarity: The aplanarity~\cite{Bjorken:1969wi} is defined as 
\begin{equation}
A=\frac{3\lambda_3}{2},
\end{equation}
and is constrained to the range $0 \leq A \leq \frac{1}{2}$.
A highly directional distribution of the energy clusters has $A=0$ and $A= 0.5$ corresponds to an isotropic distribution.
\item Fox-Wolfram Moments: The Fox-Wolfram  moments~\cite{Fox:1978vw}, $H_l$, are defined as
\begin{equation} 
H_l = \sum_{i,j} \frac{ |\vec{p}_i| |\vec{p}_j| } {E^2} \, P_l (\cos \theta_{ij}) , 
\end{equation} 	
where $\theta_{ij}$ is the opening angle between energy clusters $i$ and $j$, $E$ is the total energy of the clusters in the jet 
rest frame, the $P_l(x)$ are the Legendre polynomials.
Since the energy cluster is  a massless pseudoparticle, $H_0=1$.  For a jet that has 
a structure of two back-to-back subjets in its rest frame, $H_1=0$, $H_l\approx 1$ for even $l$, and $H_l\approx 0$ for odd $l$.
In our application,  the ratio between the second-order 
and zeroth-order Fox-Wolfram moments, $R_2$, is used as the discriminating variable.
\end{itemize}
The distributions of the jet shape variables are shown in Figure~\ref{fig:shape} for $W$ jet signal and QCD jet background.
The shape variables of the $W$ jet signal show clearly a back-to-back two body topology, while those of the
QCD jets indicate an isotropic-like distribution. They are very similar to the distributions of event shape variables observed
in $e^+e^-\to\Upsilon(4S)$~\cite{babar_book} that is  at a much lower mass scale than the $W$ boson.
This indicates that the newly introduced shape variables in the jet rest frame indeed encapsulate properties of the jet substructure
and are relatively independent of the particle mass scale. 

\begin{figure*}[p]
\begin{center}
\includegraphics[width=0.38\textwidth]{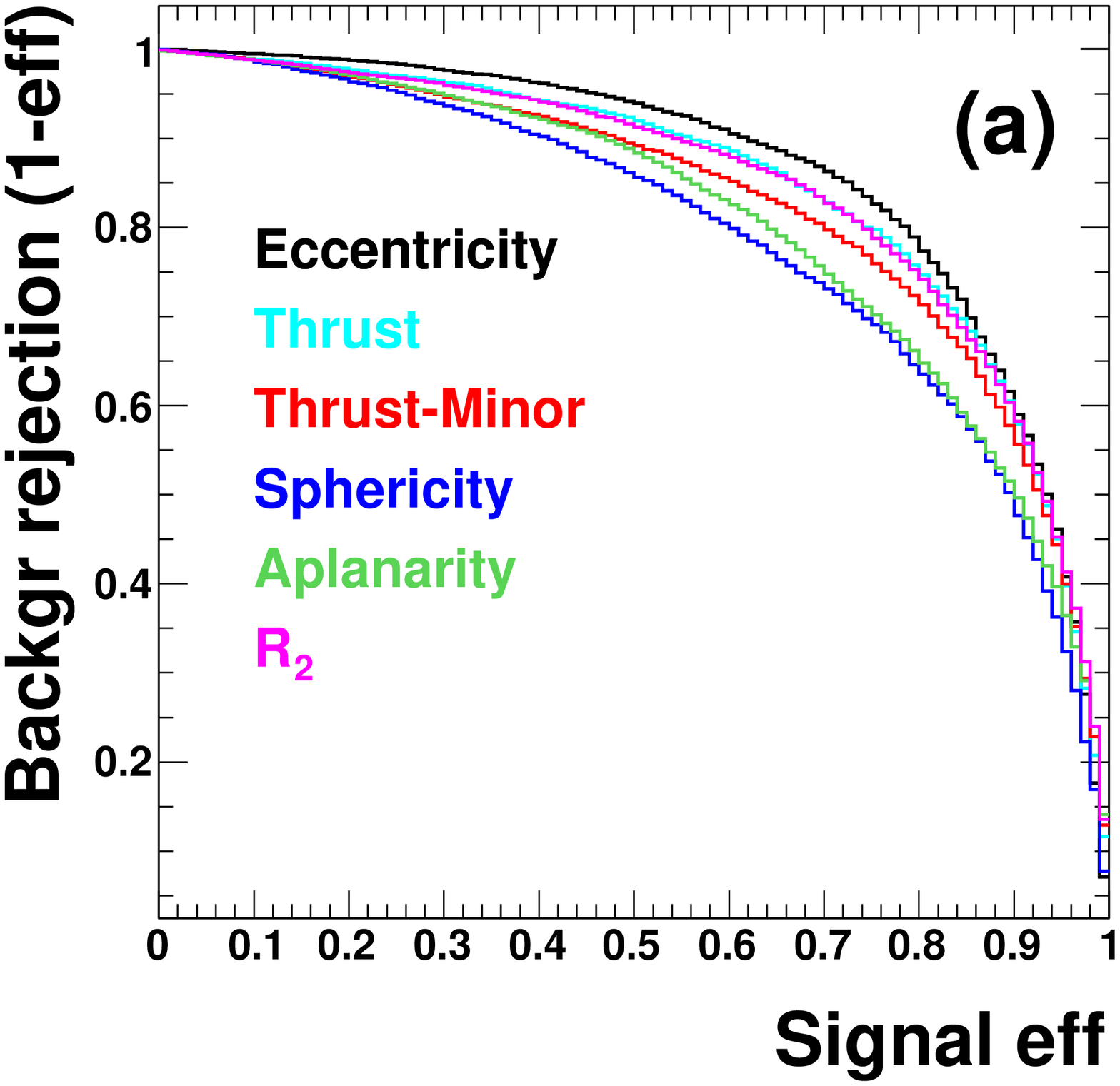}
\includegraphics[width=0.38\textwidth]{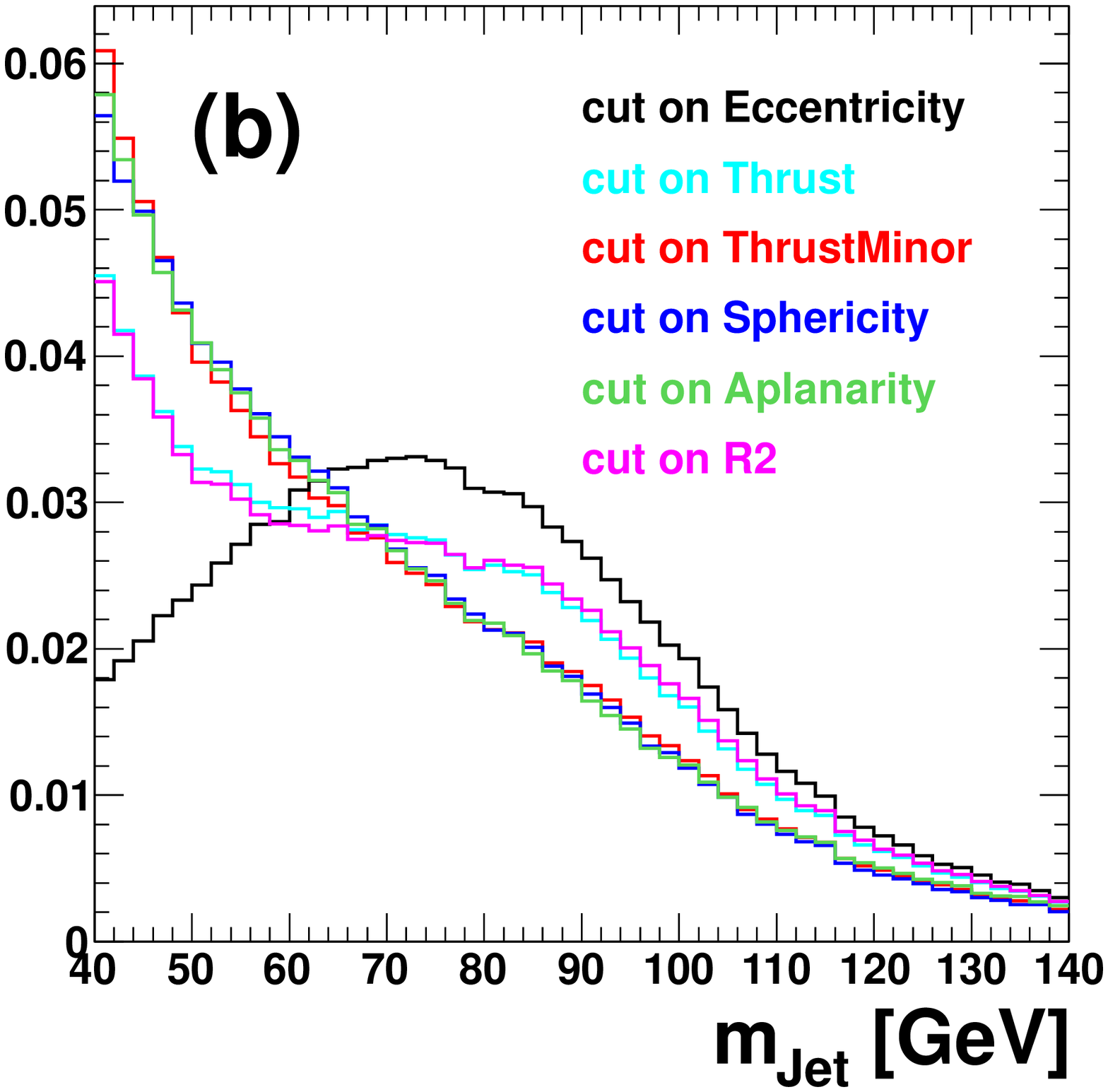}
\caption{(a) The signal efficiency of $W$ jets vs. the background rejection of QCD jets
for jet shape variables in the mass window $40\le m_{\rm jet} \le 140\gev$.
(b) The invariant mass distributions of  QCD jets after $90\%$ of them are rejected by a selection 
requirement based solely on one of the shape varialbles: Thrust, Thrust-minor, Sphericity, Aplanarity, $R_2$ and Eccentricity.
 All the distributions are normalized to unity.}
\label{fig:plotmJetvsCut}
\vspace{0.5cm}
\includegraphics[width=0.38\textwidth]{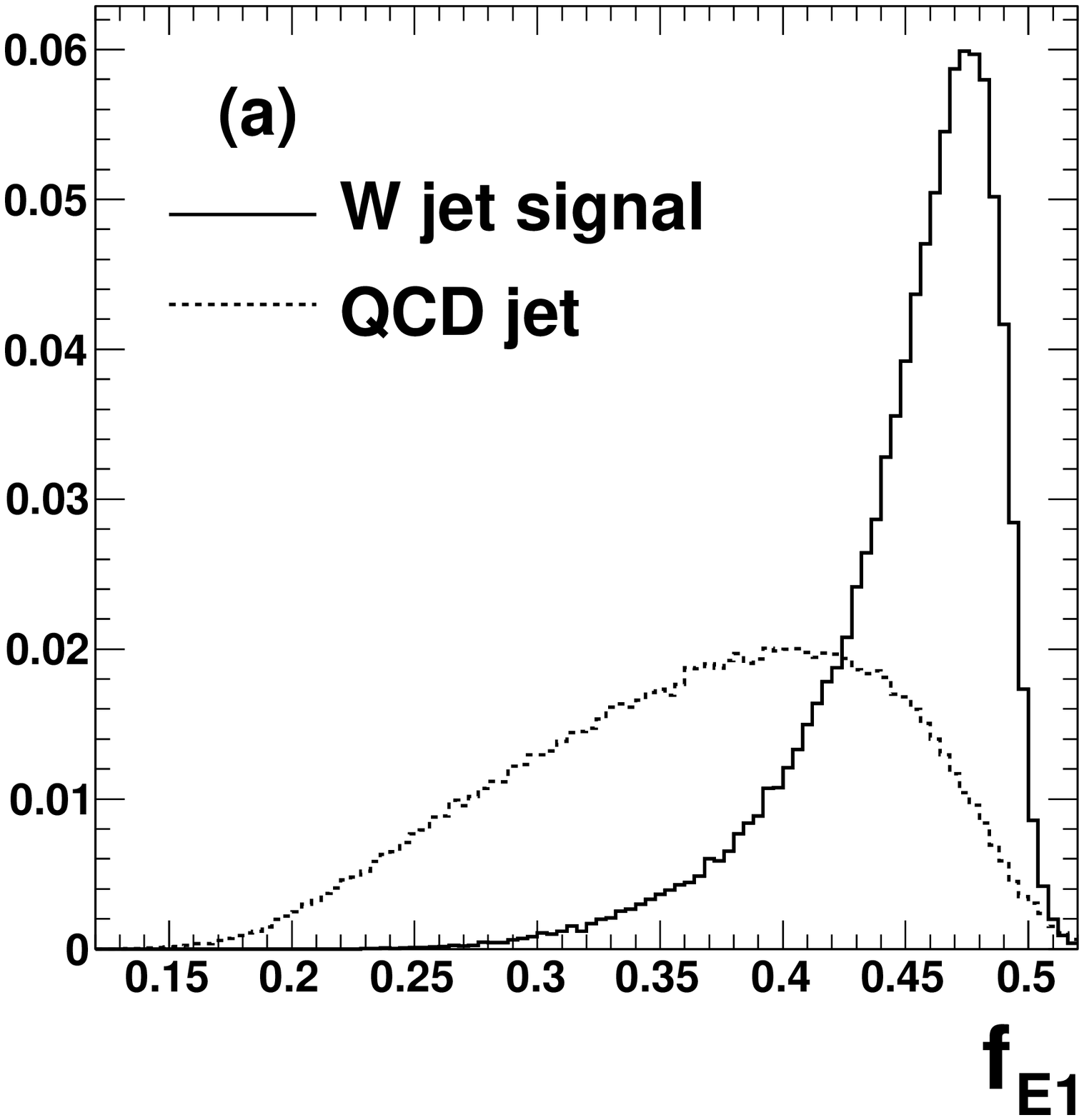}
\includegraphics[width=0.38\textwidth]{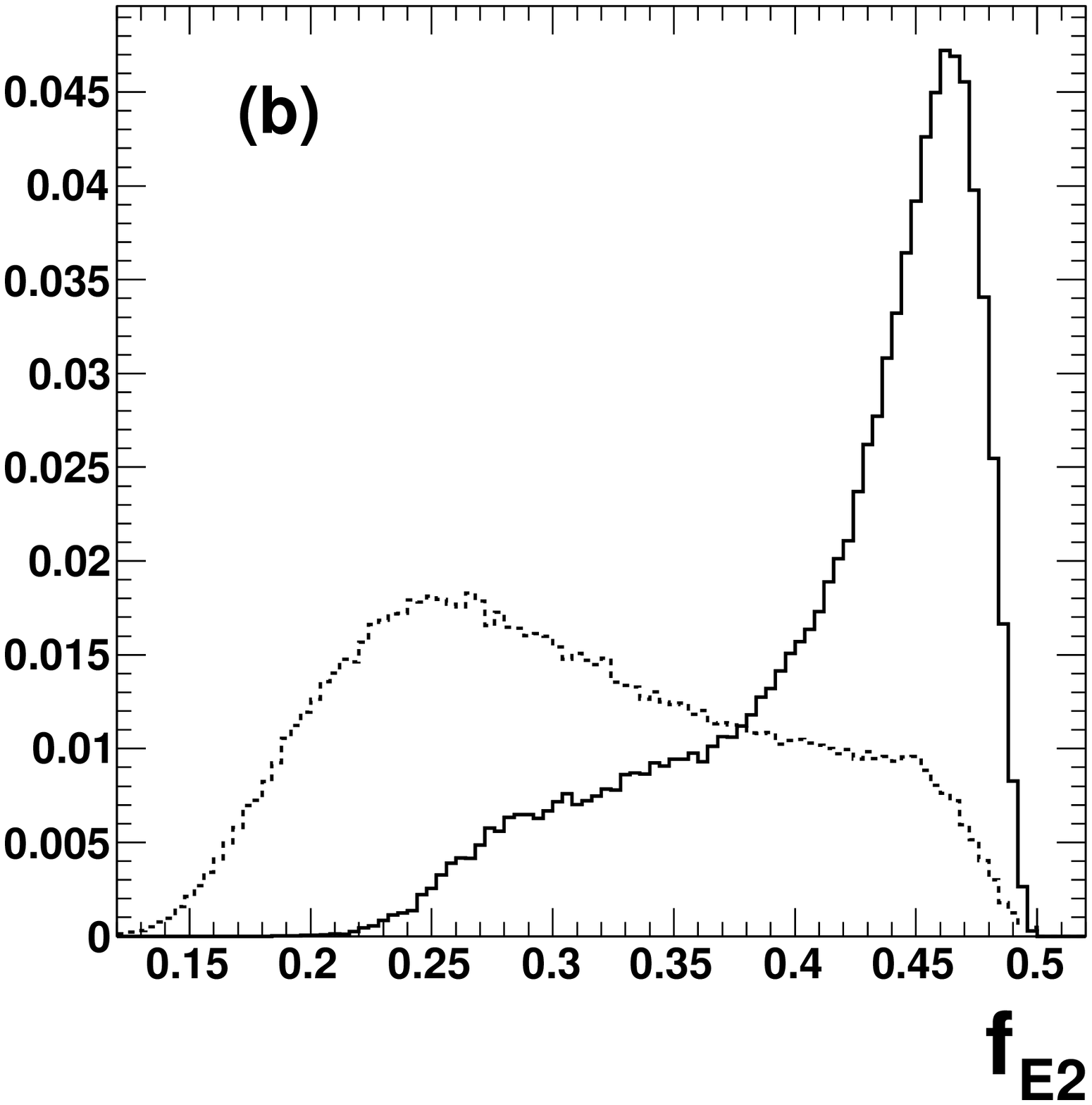}
\includegraphics[width=0.38\textwidth]{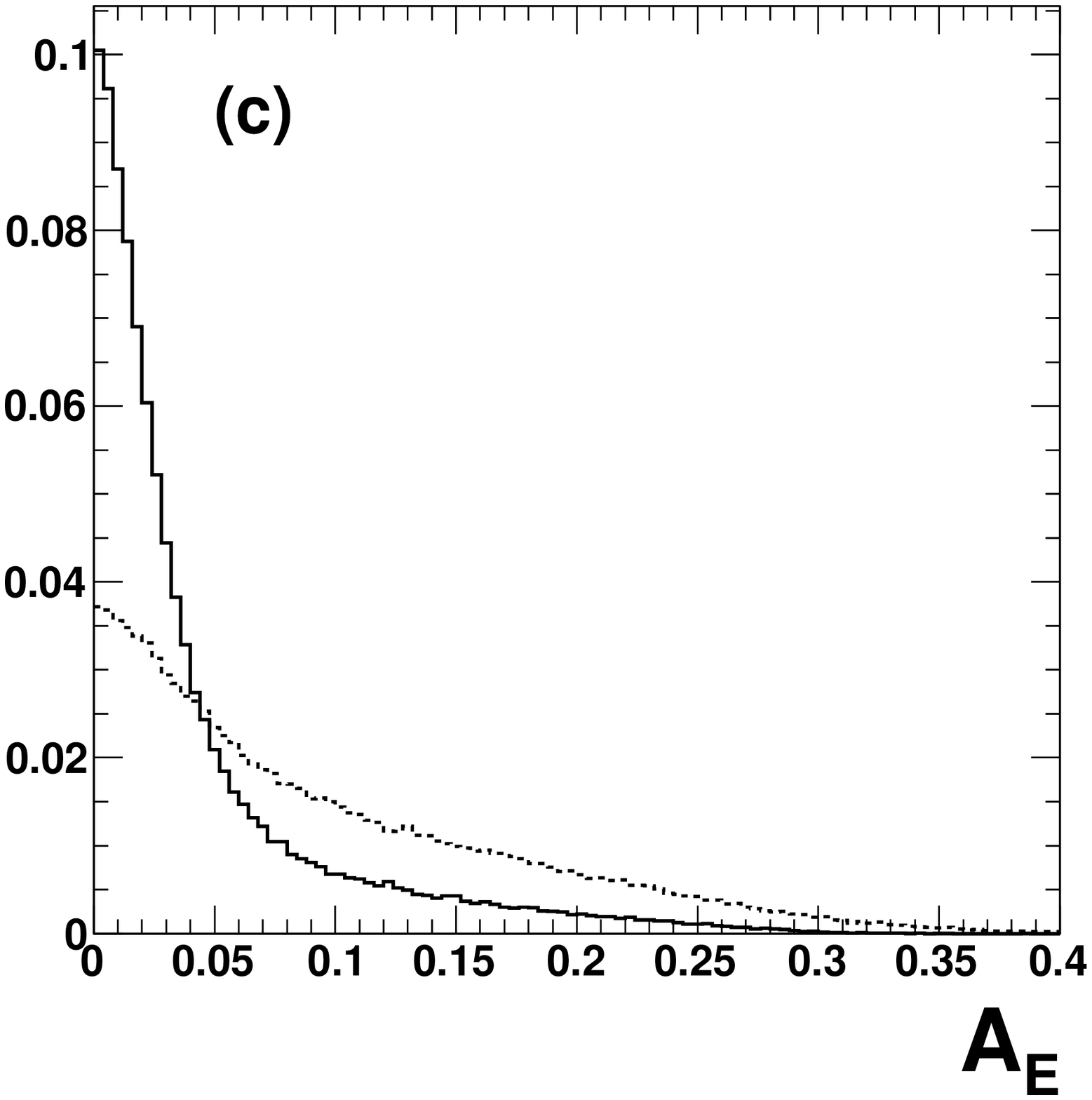}
\includegraphics[width=0.38\textwidth]{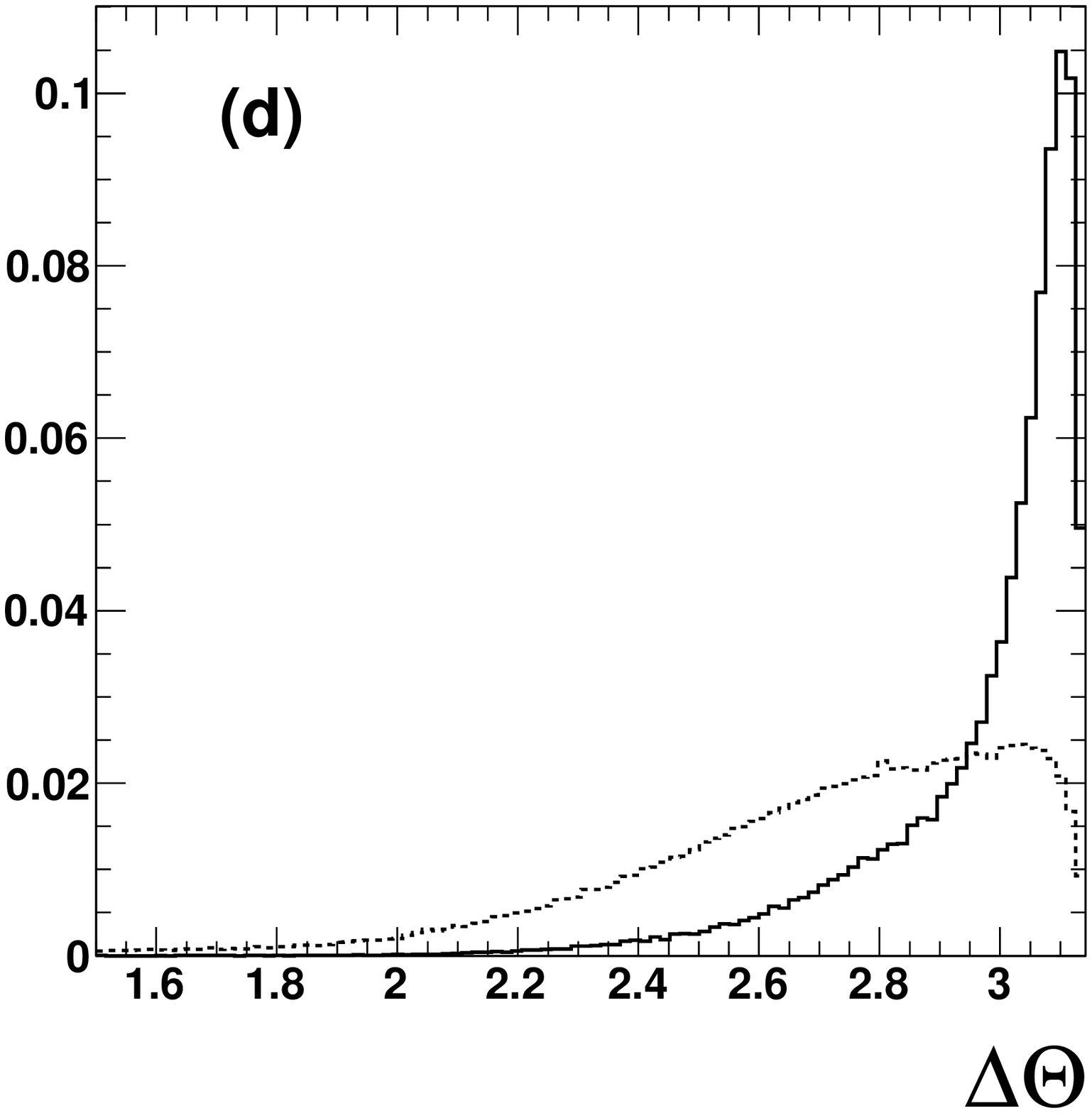}
\caption{The distributions of the kinematics of the reconstructed subjets from reclustering in the jet rest frame: 
(a) fraction of energy carried by the first leading jet, (b) fraction of energy carried by the
second leading jet, (c)  the asymmetry of the energies carried by the first and second leading jets and 
(d) the opening angle between the two leading jets. All the distributions are normalized to unity.}
\label{fig:shape_reculster}
\end{center}
\end{figure*}

We compare our new shape variables to eccentricity~\cite{Chekanov:2010vc}. The jet eccentricity is a commonly used jet  shape variable  
in the lab frame and  is defined as the difference between the maximum and minimum value of variances of jet constituents 
along the principal and minor axis, respectively. The distribution of the eccentricity  is
 also shown in Figure~\ref{fig:shape}. The new shape variables have comparable but slightly less background rejection
power while keeping the same signal efficiencies in the large mass window $40\le m_{\rm jet} \le 140\gev$,
as shown in Figure~\ref{fig:plotmJetvsCut}. However,
studies show that a variable calculated in the jet rest frame has less correlation with the jet mass
in the QCD background sample. As shown in Figure~\ref{fig:plotmJetvsCut}, 
when we tighten the selection of shape variables to reject a large fraction of the QCD background,
the eccentricity tends to reject more background events with  low $m_{\rm jet}$ and thus creates a significant kinematic enhancement
near the $W$ boson mass peak; this is not the case for some of the shape variables in the jet rest frame, such as thrust-minor,  sphericity and aplanarity. 
While for the thrust, and $R_2$, the kinematic enhancement is less significant as that for eccentricity. Therefore, our proposed jet
substructure method in the jet rest frame has an experimental advantage to separate the boosted $W/Z$ bosons from the
large QCD background.

\subsection{Reclustering}
Another important application of our proposed method is to recluster the energy clusters
of a jet to reconstruct subjets in the jet rest frame. We perform such a study using the 
Cambridge-Aachen (CA) sequential jet reconstruction algorithms with a modified distance parameter 
of $\Delta\theta = 0.6$, where $\theta$ is defined as the angle between two pseudoparticles in the jet rest frame.
We introduce several other discriminating variables:  the fraction of energy carried by the
first leading subjet ($f_{E1}$), the fraction of energy carried by the second leading subjet ($f_{E2}$), the asymmetry of the 
energy ($A_E$) that is defined as $A_E =(f_{E1}-f_{E2})/(f_{E1}+f_{E2})$, and the opening angle ($\Delta\Theta$)
between the two leading subjets.  Their distributions are shown in Figure~\ref{fig:shape_reculster}. 
Studies show that most $W$ jets have back-to-back subjets whose energies are 
around half of the $W$ boson mass, while those distributions from QCD jets are irregular. 
\begin{figure*}[!htb]
\begin{center}
\includegraphics[width=0.38\textwidth]{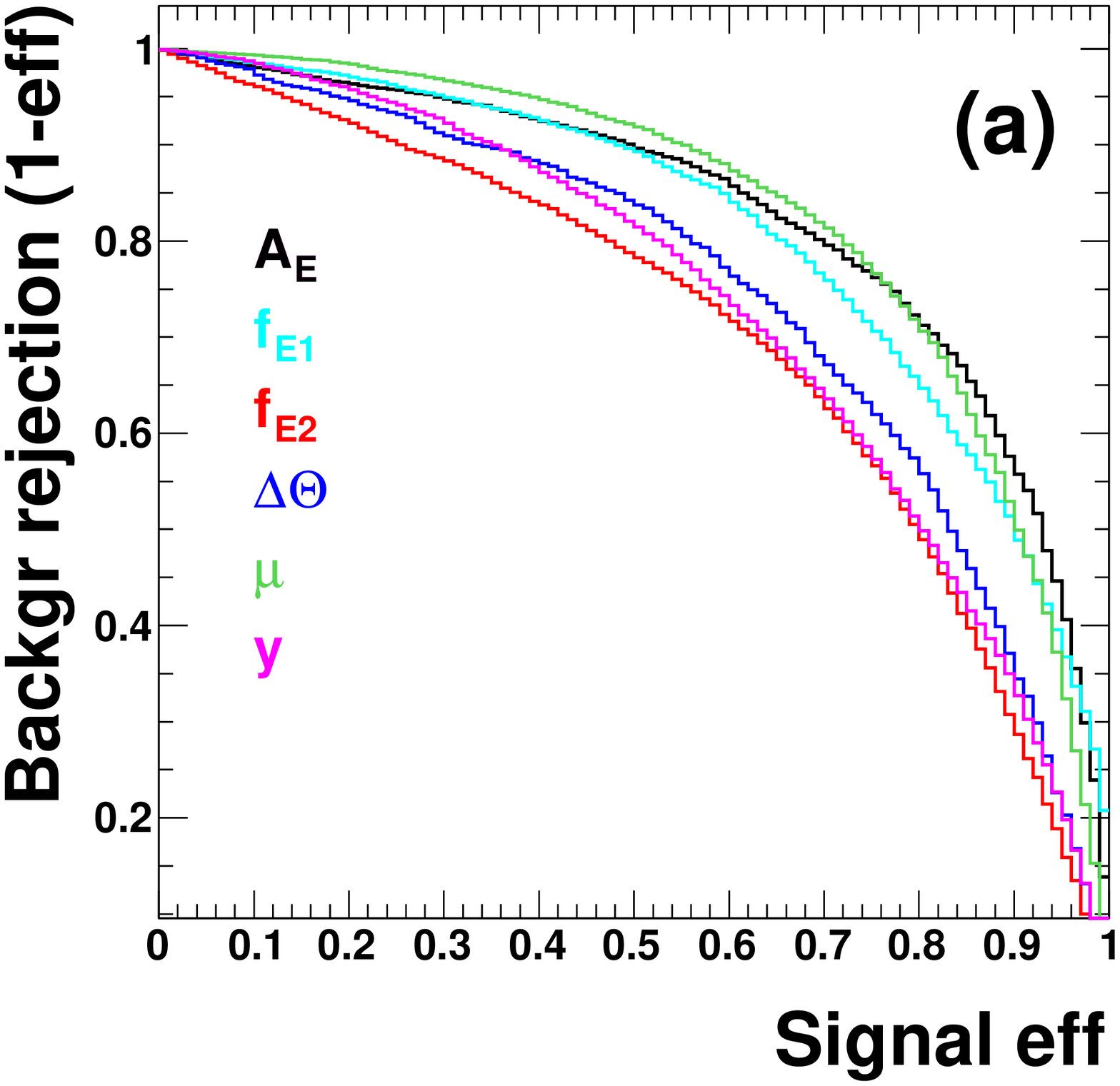}
\includegraphics[width=0.38\textwidth]{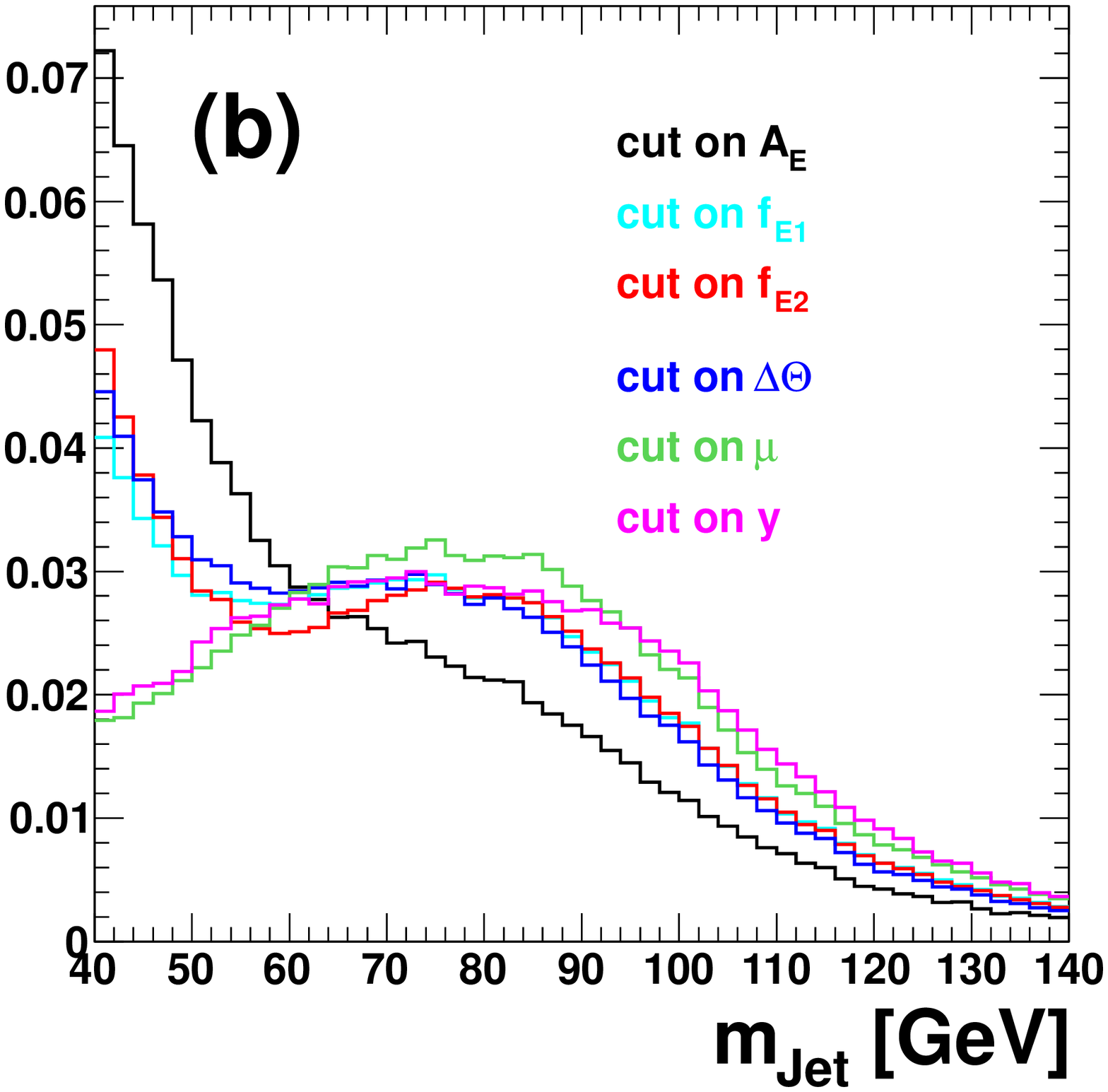}
\caption{(a) The signal efficiency of $W$ jets vs. the background rejection of QCD jets
for jet substructure variables in the mass window $40\le m_{\rm jet} \le 140\gev$.
(b) The invariant mass distributions of  QCD jets after $90\%$ of them are rejected by a selection 
requirement based solely on one of the shape variables: $f_{E1}$, $f_{E2}$, $A_E$, $\Delta\Theta$,
$\mu$ and $y$.  All the distributions are normalized to unity.}
\label{fig:plotmJetvsCutCM}
\end{center}
\end{figure*}

We compare the subjet information in the jet rest frame to the mass drop $\mu$ and splitting $y$,
the two commonly used variables by the existing two-body subjet methods, such
as YSplitter~\cite{Butterworth:2002tt} and mass-drop tagger~\cite{Butterworth:2008iy}.  
We first recluster the reconstructed jets using the $k_T$ algorithm. The last 
step of the clustering is then undone: $j\to j_1,j_2$, with $m_{j_1}>m_{j_2}$.
The mass drop and splitting are defined as $\mu\equiv m_{j_1}/m_{j}$,
and $y\equiv \min(p^2_{T,j_1},p^2_{T,j_2})\Delta R^2_{j_1j_2}/m^2_j$.
As shown in Figure~\ref{fig:plotmJetvsCutCM}, the variables  constructed using subjets in the jet rest frames
have similar background rejection power while keeping the same signal efficiencies.
We further study the correlation between the jet mass and  the variables  constructed using subjets 
and find that the traditional variables $\mu$ and $y$ tend to reject more QCD jets with the small $m_{\rm jet}$
and thus creates a significant kinematic enhancement near the $W$ boson mass peak; this is also the case for 
$f_{E1}$, $f_{E2}$ and $\Delta\Theta$, although their enhancements are relatively smaller. On the other hand, no such enhancement
is observed for the variable $A_E$, as shown in Figure~\ref{fig:plotmJetvsCutCM}. 
We  find that the variables $f_{E1}$, $f_{E2}$ and $\Delta \Theta$ 
 are  highly correlated with the shape variables we introduced before. However, the correlations between $A_E$ and 
 those variables are fairly small (less than $20\%$ ). Thus we can add 
additional discriminating power by combining them using multivariable analysis techniques, such as neural networks, boosted
decision trees, etc. 

We also point out that the rest frame subjet algorithm is infrared and collinear safe if an infrared
and collinear safe jet algorithm is used for the rest frame subjet clustering. All the sophisticated jet-grooming 
algorithms introduced in the lab frame, such as pruning~\cite{Ellis:2009su,Ellis:2009me} and trimming~\cite{Krohn:2009th},
can be easily incorporated. The leading subjets in the jet
rest frame are not much affected by the underlying event and pileup. We repeat our studies by generating MC events with 
different average numbers of multiple interactions and observe no significant difference
in the performance of the jet substructure in the center-of-mass frame. 

\section{Application}
\label{sec:app}
\begin{figure*}[!htb]
\begin{center}
\includegraphics[width=0.8\textwidth]{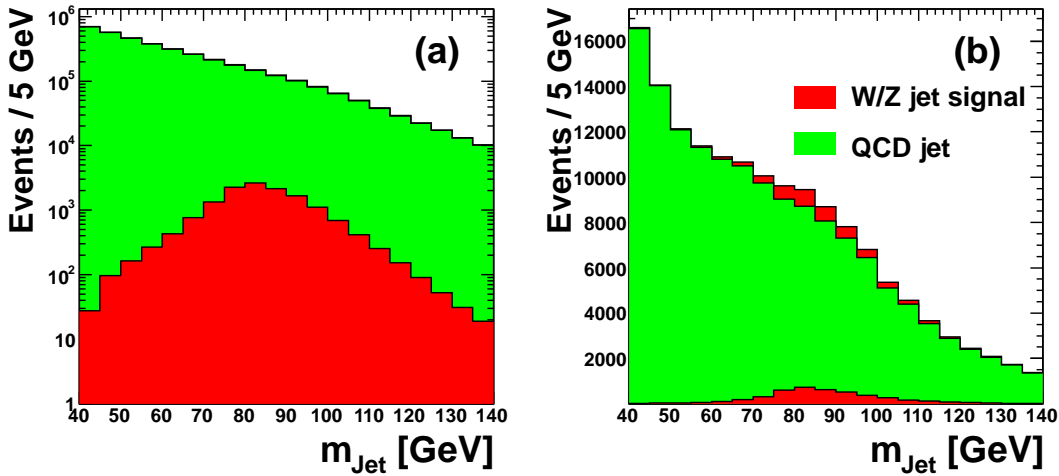}
\caption{Invariant mass of the $W/Z$ jet candidates in the MC simulated event sample that is 
equivalent to $5\,\ifb$ of LHC data at $7\,\tev$ center-of-mass energy:
(a) before the likelihood cut. (b) after the likelihood cut.}
\label{fig:mJet_inclusiveWZ}
\end{center}
\end{figure*}

As a first step, we consider the possibility of identifying boosted $W/Z$ bosons in current LHC
data.  In order to suppress the large QCD background, we construct a likelihood variable using the 
shape variables in the jet rest frame:  Thrust-Minor, Sphericity,
Aplanarity  and $A_E$. We optimize the selection cut on the likelihood variable by maximizing 
$S/\sqrt{B}$, where $S$ and $B$ are the number of the signal and background events in
the signal mass window $50\le m_{\rm jet} \le 115\gev$. The jet mass distributions
of the $W/Z$+jet and QCD jet in MC event samples are shown in Figure~\ref{fig:mJet_inclusiveWZ}.
After applying  the optimized selection cut on the likelihood, we reject more than $95\%$ of the QCD background while
keeping approximately $30\%$ of the signal. The significance of $S/\sqrt{B}$ in the signal window is more than 13.
Notice that here we treat both boosted $W$ and $Z$ bosons as signal because the jet mass resolution is 
larger than their mass difference. With enough data, their individual contributions can be extracted by fitting the signal mass distribution. 
\begin{figure*}[!htb]
\begin{center}
\includegraphics[width=0.76\textwidth]{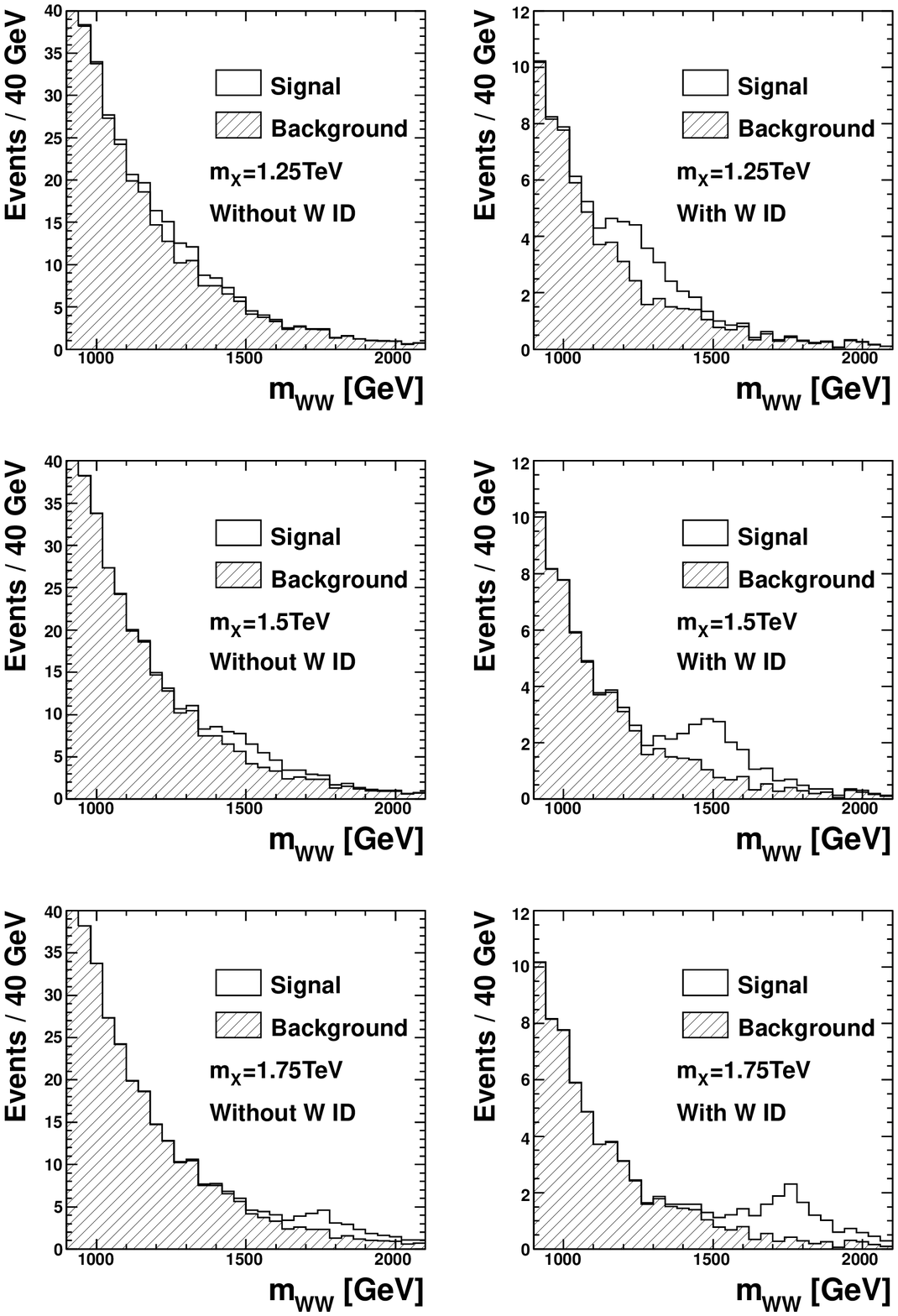}
\caption{Invariant mass of the $X\to WW$ candidates in the MC simulated event 
sample that is equivalent to  $5\ifb$ of LHC data at $7\,\tev$ center-of-mass energy.
Here we assume that the product of the production cross section of $X$ and the branching 
fraction for its decay into $WW$ pair is $50\,\rm fb$. }
\label{fig:mX_WW}
\end{center}
\end{figure*}

While our study is based on MC simulated events and the results could be somewhat optimistic, 
we point out that we have not yet used 
all the available jet substructure variables.  More sophisticated multivariable analysis techniques such as neural network
or boosted decision trees will further compensate for any potential underestimate of the background. As a result, we expect
to establish the signal of hadronically decaying $W/Z$ jets and measure their  inclusive production cross section 
with current LHC data. Such a measurement is not only a prerequisite of any NP search using boosted $W/Z$ bosons,
but also a model-independent test of the SM. Any excess of boosted $W$ and $Z$ bosons will be a promising
hint of the existence of NP. 

Reconstructed  decays of $W$'s and $Z$'s to jets 
can be used to search for NP with specific final state signatures. Here we
demonstrate such applications by considering  a heavy resonance that decays to a $WW$ final state:
$pp\to X\to WW$, where the $X$ is a new heavy resonance beyond the SM, such as a 
new heavy gauge boson, or a Kaluza-Klein $Z^\prime$ in the Randall-Sundrum (RS) model, etc. 
\begin{figure}[!htb]
\begin{center}
\includegraphics[width=0.48\textwidth]{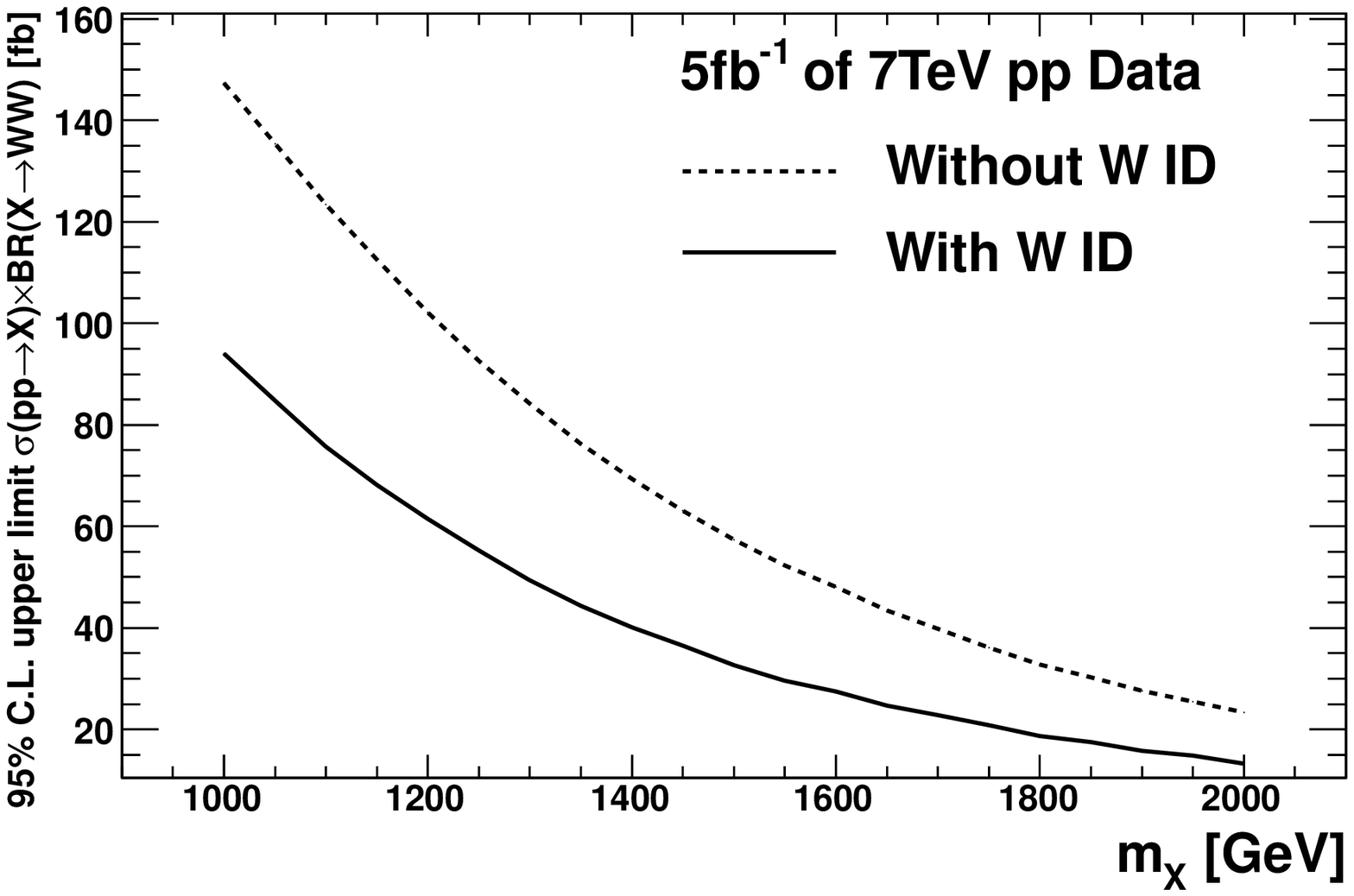}
\caption{The expected 95\% C.L. upper limit on the product of the production cross section of a heavy resonance $X$
and the branching fraction for its decay into a $WW$ final state, as a function of assumed $X$ mass. The solid black curve represent
the upper limit when we apply the explicit $W$  jet identification, the dashed black curve shows the upper limit without
using $W$  jet identification. Here we assume a width of $125\gev$ of the heavy resonance and
 $5\ifb$ of LHC data at $7\,\tev$ center-of-mass energy. }
\label{fig:limit_XtoWW}
\end{center}
\end{figure}
We consider a search for an $X$ signal by fully reconstructing the $X$ signal candidate in the
decay mode where one $W$ boson decays leptonically  and the other one decays hadronically.
Note that for such high mass  ($\approx1\tev$) resonance decays, more than $90\%$ of the events have the two quarks 
from the $W$ boson decay within a cone of $\Delta R<0.4$. This makes it very difficult to identify two separate jets.
As a result,  we select the leading jet with $p_{\rm T}>300\gev$ and $|\eta|<2.5$ in an event as the 
hadronically decaying $W$ boson candidate. The jet is reconstructed with the anti-$k_{\rm T}$ algorithm with a distance
parameter of $\Delta R=0.6$. 
The leptonically-decaying $W$ boson is reconstructed by requiring one isolated lepton with $p_{\rm T}>20\gev$, $|\eta|<2.5$
and more than $25\gev$ of missing transverse energy in the event. The presence of only one neutrino in the final state allows for the
reconstruction of its momentum by requiring transverse momentum conservation and applying the $W$ boson mass
constraint. In doing so, we obtain two solutions of the neutrino $p_z$, which leads to two reconstructed $WW$ masses.
Studies show that the difference between the two reconstructed masses  is small so we take the minimum of the two as the reconstructed 
mass of the resonance $X$ ($m_{X}$). The major SM backgrounds are the production of $W$+jets, $WW$, and $t\bar{t}$.
In order to reduce the background, we  explicitly identify the boosted $W$ jets by requiring their jet mass to be within $20\gev$
of the $W$ boson mass, which is slightly more than twice that of the expected $W$ jet mass resolution in the MC simulation.  
We also apply $W$ jet identification ($W$ ID), a selection on the likelihood variable as described before, to reject
more than half of the QCD jets while keeping more than 80\% of the signal. 
The invariant mass distributions of the $X\to WW$ candidates in the MC simulated event 
sample that is equivalent to  $5\ifb$ of LHC data at $7\,\tev$ center-of-mass energy are shown in Figure~\ref{fig:mX_WW}.
We estimate the expected 95\% C.L. upper limit on the product of the
production cross section of a heavy resonance $X$ and a branching fraction for its decay into $WW$ pair. The expected limit for
$5\ifb$ of LHC data at $7\tev$ center-of-mass energy is plotted 
as a function of the assumed $X$ mass, as shown in Figure~\ref{fig:limit_XtoWW}. For comparison, we also plot the 
expected 95\% upper limit without explicit $W$ jet identification. It is clear that the boosted $W$ jet technique can
significantly improve our experimental sensitivity.
The above discussion can be directly applied to searches for heavy resonances that decay to other di-boson final
states, such as $X\to ZZ/WZ$. 

\section{Conclusion}
In this paper we introduce a new approach to study jet substructure in the center-of-mass frame of the jet. 
We demonstrate that it can be used to discriminate boosted heavy particles from  QCD jets.
The method suggested in this paper is a proof of concept and is complementary to the existing algorithms to identify boosted
heavy particles based on jet substructure. 
\label{sec:conclusion}

\section{Acknowledgments}
We thank Sergei Chekanov, Jim Cochran, Nils Krumnack, Soeren Prell and German Valencia 
for many discussions and valuable comments on the manuscript. This work is
supported by the US ATLAS fellowship.

\end{document}